%% file: main.tex
\documentclass[copyright, creativecommons]{eptcs}
\providecommand{\event}{PDMC 2009} 
\providecommand{\volume}{??}
\providecommand{\anno}{20??}
\providecommand{\firstpage}{1}
\providecommand{\eid}{??}

\usepackage{amsmath,amsfonts,amssymb}

\usepackage{float}
\usepackage{xspace}
\usepackage{epsfig}
\usepackage{algorithm, algorithmic}
\usepackage{subfigure}
\usepackage{multicol}

\newcommand{\BA}{\mathit{BA}}

\newcommand{\probstmt}{2.1\xspace}
\newif\ifthink\thinkfalse
\newif\iffailstop\failstopfalse

\newcommand{\thr}{\mathit{th}}

\newcommand{\pp}{\mathit{prt}}

\newcommand{\ds}{\mathit{ds}}

\newcommand{\tr}{\mathit{tr}}
\newcommand{\gtr}{\mathit{gtr}}

\newcommand{\id}{\mathit{id}}
\newcommand{\gAll}{\mathit{gAll}}

\title{ Parallelizing Deadlock Resolution in Symbolic Synthesis of Distributed Programs\thanks{This work was partially sponsored by the COMBEST European project, NSF CNS 0914913 , and ONR Grant N00014-01-1-0744.}}
\author{Fuad Abujarad
\institute{Department of Computer Science and Engineering\\Michigan State University\\East Lansing, MI 48824, USA}
\email{abujarad@cse.msu.edu}
\and
Borzoo Bonakdarpour
\institute{VERIMAG\\Centre \'{E}quation, 2 ave de Vignate\\38610 Gi\`{e}res, France}
\email{borzoo@imag.fr}
\and
Sandeep S. Kulkarni
\institute{Department of Computer Science and Engineering\\Michigan State University\\East Lansing, MI 48824, USA}
\email{sandeep@cse.msu.edu}
}
\def\titlerunning{Parallelizing Deadlock Resolution in Symbolic Synthesis of Distributed Programs}
\def\authorrunning{F. Abujarad, B. Bonakdarpour \&  S. S. Kulkarni}

\begin{document}
\maketitle

\input{macro.tex}

\input{abs}

\input{intro}

\input{pr}

\input{agreementproblem}
\input{readwrite}

\input{DL}

\input{FG}

\input{results}

\input {analyses}

\input{CGnew}

\input{related}

\input{concl}

{
\bibliographystyle{eptcs}
\bibliography{bibliography}}
\end{document}

%% file: macro.tex
\newcommand{\ad}{\wedge}
\newcommand{\intersection}{\cap}
\newcommand{\orr}{\vee}
\newcommand{\bx}{[]}

\newcommand{\ms}{\mathit{ms}}
\newcommand{\ns}{\mathit{ns}}
\newcommand{\mt}{\mathit{mt}}
\newcommand{\MT}{\mathit{MT}}
\newcommand{\MS}{\mathit{MS}}
\newcommand{\rank}{\mathrm{Rank}}
\newcommand{\SP}{\mathrm{SP}}
\newcommand{\tru}{\mathit{true}}
\newcommand{\fals}{\mathit{false}}
\newcommand{\xarrow}[1]{\xrightarrow{#1}}
\newcommand{\guard}{\mathit{g}}
\newcommand{\group}{\mathit{Group}}
\newcommand{\statement}{\mathit{st}}
\newcommand{\zplus}{\mathbb{Z}_{\geq 0}}
\newcommand{\zpluszero}{\mathbb{Z}_{> 0}}
\newcommand{\rplus}{\mathbb{R}_{\geq 0}}
\newcommand{\spec}[1]{\mathit{SPEC}_{#1}}
\newcommand{\sspec}[1]{\mathit{SSPEC}_{#1}}
\newcommand{\comput}{\overline{\sigma}}
\newcommand{\prefix}{\overline{\alpha}}
\newcommand{\suffix}{\overline{\beta}}
\newcommand{\Bullet}{\noindent \hspace*{4mm}\ensuremath{\bullet\;\;}}
\newcommand{\notate}{\noindent \emph{Notation. } \ }
\newcommand{\infcomput}{(\sigma_0, \tau_0) \rightarrow (\sigma_1, \tau_1) \rightarrow \cdots}
\newcommand{\infcomputp}{(\sigma'_0, \tau'_0) \rightarrow (\sigma'_1, \tau'_1) \rightarrow \cdots}
\newcommand{\fcomput}{(\sigma_0, \tau_0) \rightarrow (\sigma_1, \tau_1) \rightarrow \cdots (\sigma_n, \tau_n)}
\newcommand{\BR}[2]{P_{#1} \mapsto_{\leq #2_{#1}} Q_{#1}}
\newcommand{\fstar}[2]{(#1)^*_{#2}}
\newcommand{\exmp}{\noindent \textbf{Example. } \ }
\newcommand{\atime}[1]{\mathit{time}(#1)}
\newcommand{\pprime}{\ensuremath{\prog'}\xspace}
\newcommand{\gprime}{\ensuremath{g'}\xspace}
\newcommand{\cprime}{\ensuremath{c'}\xspace}
\newcommand{\stprime}{\ensuremath{st'}\xspace}
\newcommand{\acprime}{\ensuremath{ac'}\xspace}
\newenvironment{defin}[1]{\vspace{-2mm}\begin{definition}{(\bf #1) \ }}{\end{definition}\vspace{-2mm}}

\newcommand{\algorithmicinput}{\textbf{Input:}}
\newcommand{\INPUT}{\item[\algorithmicinput]}
\newcommand{\algorithmicoutput}{\textbf{Output:}}
\newcommand{\OUTPUT}{\item[\algorithmicoutput]}

%% file: abs.tex
\begin{abstract}

Previous work has shown that there are two major complexity barriers in the synthesis of fault-tolerant distributed programs: (1) generation of fault-span, the set of states reachable in the presence of faults, and (2) resolving deadlock states, from where the program has no outgoing transitions. Of these, the former closely resembles with model checking and, hence, techniques for efficient verification are directly applicable to it. Hence, we focus on expediting the latter with the use of multi-core technology. 

We present two approaches for parallelization by considering different design choices. The first approach is based on the computation of equivalence classes of program transitions (called \textit{group} computation) that are needed due to the issue of distribution (i.e., inability of processes to atomically read and write all program variables). We show that in most cases the speedup of this approach is close to the ideal speedup and in some cases it is superlinear. 
The second approach uses traditional technique of partitioning deadlock states among multiple threads. However, our experiments show that the speedup for this approach is small. Consequently, our analysis demonstrates that a simple approach of parallelizing the group computation is likely to be the effective method for using multi-core computing in the context of deadlock resolution.

\end{abstract}

\vspace*{0mm}
\begin{center}
\begin{minipage}[t]{\columnwidth} \textbf{Keywords: Program transformation, Symbolic synthesis, Multi-core algorithm, Distributed programs.}
\end{minipage}

\end{center}

%% file: intro.tex
\section{Introduction}
\label{sec: intro}

Given the current trend in processor design where the number of transistors keeps growing as directed by Moore's law, but where clock speed remains relatively flat, it is expected that multi-core computing will be the key for utilizing such computers most effectively. As argued in \cite{Maurice08}, it is expected that programs and protocols from distributed computing will be especially beneficial in exploiting such multi-core computers.

One of the crucial issues in distributed computing is fault-tolerance. Moreover, as part of maintenance, it may be necessary to modify a program to add fault-tolerance to faults that were not considered in the original design. In such maintenance, it would be required that the existing functional properties of the program continue to be preserved during the addition of fault-tolerance, i.e., no bugs should be introduced in such addition. For this reason, it would be highly beneficial if one could add such fault-tolerance properties using automated techniques. 

One difficulty in adding fault-tolerance using automated techniques, however, is its complexity. In our previous work \cite{bk07}, we developed a symbolic (BDD-based) algorithm for adding fault-tolerance to distributed programs specified in terms of transition system with state space larger than $10^{30}$. We also identified a set of bottlenecks that compromise the effectiveness of our algorithm. Based on the analysis of the experimental results from \cite{bk07}, we observed that depending upon the structure of the given distributed intolerant program, performance of synthesis suffers from two major complexity obstacles, namely \emph{generation of fault-span} (i.e., the set of reachable states in the presence of faults) and \emph{resolution of deadlock states}.

Our focus in this paper is to evaluate effectiveness of different approaches that utilize multi-core computing to reduce the time complexity of adding fault-tolerance to distributed programs. In particular, we focus on the second problem, i.e., resolution of deadlock states. Deadlock resolution is especially crucial in the context of dependable systems, as it guarantees that the synthesized fault-tolerant program meets its liveness requirements even in the presence of faults. A program may reach a deadlock state due to the fact that faults perturb program to a new state that was not considered in the fault-intolerant program. Or, it may reach a deadlock state, as some program actions are removed (e.g., because they violate safety in the presence of faults). To resolve a deadlock state, we either need to provide {\em recovery actions} that allow program to continue its execution or \emph{eliminate} the deadlock state by preventing the program execution from reaching it. 

To evaluate the effectiveness of multi-core computing, we first need to identify bottleneck(s) where multi-core features can provide the maximum impact. To this end, we present two approaches for parallelization. The first approach is based on the distributed nature of the program being synthesized. In particular, when a new transition is added (respectively, removed), since the process executing it has only a partial view of the program variables, we need to add (respectively, remove) a \textit{group} of transitions based on the variables that cannot be read by the process. The second approach is based on partitioning deadlock states among multiple threads. We show that while in most cases the speedup of the first approach is close to the ideal speedup and in some cases it is superlinear the second approach provides a small performance benefit. Based on the analysis of these results, we argue that the simple approach that parallelizes the group computation is likely to provide maximum benefit in the context of deadlock resolution for synthesis of distributed programs. 

\noindent {\bf Contributions of the paper. } \ Our contributions in this paper is as follows:

\begin{itemize}
\item We present two approaches for expediting resolution of deadlock states in automated synthesis of fault-tolerance. 
\item We analyze these approaches in terms of three classic examples from distributed computing: Byzantine agreement \cite{lsp82}, agreement in the presence of both failstop and Byzantine faults, and token ring \cite{ak981}.

\item We discuss different design choices considered in these two approaches. 

\end{itemize}

\noindent {\bf Organization of the paper. } \ The rest of the paper is organized as follows. In Section \ref{sec:program}, we define distributed programs and specifications. We illustrate the issues involved in the synthesis problem in the context of Byzantine agreement in Section \ref{sec:issues}. We present our two approaches, the corresponding experimental results and analysis in Sections \ref{sec:FG} and \ref{sec:cgp}. 
Finally, we discuss related work in Section \ref{sec:related} and conclude in Section \ref{sec:concl}.

%% file: pr.tex
\section{Programs, Specifications and Problem Statement}
\label{sec:program}

In this section, we define the problem statement for adding fault-tolerance. We begin with a fault-intolerant program, say $p$, that is correct in the absence of faults. We let $p$ be specified in terms of its state space, $S_p$, and a set of transitions, $\delta_p \subseteq S_p \times S_p$. Whenever it is clear from the context, we use $p$ and its transitions $\delta_p$ interchangeably. A sequence of states, $\langle s_0, s_1, ...\rangle$ (denoted by $\sigma$) is a computation of $p$ iff (1) $(\forall j : 0 < j < \mathit{length(\sigma)} : (s_{j-1}, s_j) \in p)$, i.e., in each {\em step} of this sequence, a transition of $p$ is executed, and (2) if the sequence is finite and terminates in $s_j$ then $\forall s' :: (s_j, s') \not \in p$ (a finite computation reaches a state from where there is no outgoing transition).  A special subset of $S_p$, say $S$, identifies an {\em invariant} of $p$. By this we mean that if a computation of $p$ begins in a state where $S$ is true, then (1) $S$ is true at all states of that computation and (2) the computation is {\em correct}. Since the algorithm for addition of fault-tolerance begins with a program that is correct in the {\em absence} of faults, we do not explicitly need the program specification in the absence of faults. Instead, the predicate $S$ is used to determine states where the fault-tolerant program could recover in the presence of faults.

The goal of an algorithm that adds fault-tolerance is to begin with a program $p$ and its invariant $S$ to derive a fault-tolerant program, say $p'$, and its invariant, say $S'$. Clearly, one additional input to such an algorithm is $f$, the class of faults to which tolerance is to be added. Faults are also specified as a subset of $S_p \times S_p$. Note that this allows modeling of different types of faults, such as transients, Byzantine (see Section \ref{sec:agreement}), crash faults, etc. Yet another input to the algorithm for adding fault-tolerance is a safety specification, say $\spec{bt}$, that should not be violated in the presence of faults. We let $\spec{bt}$ also be specified by a set of {\em bad} transitions, i.e., $\spec{bt}$ is a subset of $S_p \times S_p$\footnote{As shown in \cite{ke05TDSCbtvsbp}, permitting more general specifications can significantly increase the complexity of synthesis. We also showed that representing safety specification using a set of transitions is expressive enough for most practical programs.}. Thus, it is required that in the presence of faults, the program should not execute a transition from $\spec{bt}$. 

Now we define the problem of adding fault-tolerance. Let the input program be $p$, invariant $S$, faults $f$, and safety specification $\spec{bt}$. Since our goal is to add fault-tolerance only, we require that no new computations are added in the {\em absence} of faults. Thus, if the output after adding fault-tolerance is program $p'$ and invariant $S'$, then $S'$ should not include any states that are not in $S$; without this restriction, $p'$ can begin in a state from where the correctness of $p$ is unknown. Likewise, if $(s_0, s_1)$ is a transition of $p'$ and $s_0 \in S'$ then $(s_0, s_1)$ must also be a transition of $p$; without this restriction, $p'$ will have new computations in the absence of faults. Also, if $p'$ has no outgoing transition from state $s_0 \in S'$, then it must be the case that $p$ also has no outgoing transitions from $s_0$; without this restriction, $p'$ may deadlock in a state that had no correspondence with $p$. 

Additionally, $p'$ should be fault-tolerant. Thus, during the computation of $p'$, if faults from $f$ occur then the program may be perturbed to a state outside $S'$. Just like the invariant captured the boundary up to which the program can reach in the {\em absence} of faults, we can identify a boundary upto which the program can reach in the {\em presence} of faults. Let this boundary (denoted by fault-span) be $T$. Thus, if any transition of $p$ or $f$ begins in a state where $T$ is true, then it must terminate in a state where $T$ is true. Moreover, if $p'$ is permitted to execute for a long enough time without perturbation of a fault, then $p'$ should reach a state where its invariant $S'$ is true. Based on this discussion, we define the problem of adding fault-tolerance as follows:

 {\bf Problem statement \probstmt } \ Given $p$, $S$, $f$ and  $\spec{bt}$, identify $p'$ and $S'$ such that:

\begin{itemize}
\item $(C1)$: Constraints on the invariant
\begin{itemize}
\item $S' \neq \phi$,
\item $S' \Rightarrow S$,
\end{itemize}
\item $(C2)$: Constraints on transitions within invariant
\begin{itemize}
\item $(s_0, s_1) \in p' \ \wedge \ s_0 \in S' \ \ \ \Rightarrow \ \ \ ( (s_1 \in S')  \wedge (s_0, s_1) \in p$), 
\item  $s_0 \in S' \wedge (\forall s_1 :: (s_0, s_1) \not \in p') \ \ \ \Rightarrow \ \ \ (\forall s_1 :: (s_0, s_1) \not \in p), \ \ $ and
\end{itemize}
\item $(C3)$  There exists $T$ such that 
\begin{itemize}
\item $S' \Rightarrow T$,
\item $s_0 \in T \wedge (s_0, s_1) \in (p' \cup f) \ \  \ \Rightarrow \ \ \  s_1 \in T \wedge (s_0, s_1) \not \in \spec{bt}$
\item $s_0 \in T \wedge \langle s_0, s_1, ... \rangle$ is a computation of $p'$ $\ \ \ \Rightarrow \ \ \  (\exists j : 0 < j < \mathit{length}(\langle s_0, s_1, ... \rangle)  : s_j \in S')$
\end{itemize}
\end{itemize}

%% file: agreementproblem.tex
\section{Issues in Automated Synthesis of Fault-Tolerant Programs}
\label{sec:issues}

In this section, we use the example of \emph{Byzantine agreement} \cite{lsp82} (denoted $\BA$) to describe the issues in automated synthesis of fault-tolerant programs. Towards this end, in Section \ref{sec:agreement}, we describe the inputs used for synthesizing the Byzantine agreement problem. Subsequently, in Section \ref{sec:rw}, we identify the need for explicit modeling of read-write restrictions imposed by the nature of the distributed program. Finally, in Section \ref{sec:dl}, we describe how deadlock states get created while revising the program for adding fault-tolerance and illustrate our approach for managing them. 

\subsection{Input for Byzantine Agreement Problem}
\label{sec:agreement}

The Byzantine agreement problem ($\BA$) consists of a \emph{general}, say $g$, and three (or more) \emph{non-general} processes, say $j$, $k$, and $l$. The agreement problem requires a process to copy the decision chosen by the general ($0$ or $1$) and finalize (output) the decision (subject to some constraints). Thus, each process of $\BA$ maintains a decision $d$; for the general, the decision can be either $0$ or $1$, and for the non-general processes, the decision can be $0$, $1$, or $\bot$, where the value $\bot$ denotes that the corresponding process has not yet received the decision from the general. Each non-general process also maintains a Boolean variable $f$ that denotes whether that process has finalized its decision. For each process, a Boolean variable $b$ shows whether or not the process is Byzantine; the read/write restrictions (described in Section \ref{sec:rw}), ensure that a process cannot determine if other processes are Byzantine. A Byzantine process can output different decision to different processes. Thus, a state of the program is obtained by assigning each variable, listed below, a value from its domain. And, the state space of the program is the set of all possible states. 

\begin{tabbing}
$V =$ \= $\{d.g\} \; \cup$ \hspace{3.2cm} \= (the general decision variables):\{$0$, $1$\}\\
\> $\{d.j, d.k, d.l\} \; \cup$ \> \= (the processes decision variables):\{$0$, $1$, $\bot$\}\\
\> $\{f.j, f.k, f.l\} \; \cup$ \> (finalized?):\{$\fals$, $\tru$\}\\
\> $\{b.g, b.j, b.k, b.l\}$. \> (Byzantine?):\{$\fals$, $\tru$\}
\end{tabbing}

\noindent {\bf Fault-intolerant program. } \ 
 To concisely describe the transitions of the (fault-intolerant) version of $\BA$, we use guarded commands of the form $g \longrightarrow st$, where $g$ is a predicate involving the above program variables and $st$ updates the above program variables.  The command $g \longrightarrow st$ corresponds to the set of transitions $\{ (s_0, s_1) : g$ is true in $s_0$ and $s_1$ is obtained by {\em executing} $st$ in state $s_0 \}$. Thus, the transitions of a non-general process, say $j$, is specified by the following two actions:

{
\small
\begin{tabbing}
\hspace*{.5cm}$BA_{\mathit{intol}_{j}}::$
 \= $\BA1_j \; :: \; (d.j = \bot) \; \ad \; (f.j = \fals) \; \ad \; (b.j = \fals)$ \hspace*{1mm} \= $\longrightarrow$ \hspace*{1mm} \= $d.j := d.g$\\
\> $\BA2_j \; :: \; (d.j \neq \bot) \; \ad \; (f.j = \fals)  \; \ad \; (b.j = \fals)$ \> $\longrightarrow$ \> $f.j := \tru$
\end{tabbing}
}

\noindent We include similar transitions for $k$ and $l$ as well. Note that the general does not need explicit actions; the action by which the general sends the decision to $j$ is modeled by $\BA1_j$. 

\noindent {\bf Specification. } \ 
The safety specification of $\BA$ requires \emph{validity} and \emph{agreement}. \emph{Validity} requires that if the general is non-Byzantine, then the final decision of a non-Byzantine, non-general must be the same as that of the general. Additionally, \emph{agreement} requires that the final decision of any two non-Byzantine, non-generals must be equal. Finally, once a non-Byzantine process finalizes (outputs) its decision, it cannot change it.

\noindent {\bf Faults. } \ 
A fault transition can cause a process to become Byzantine, if no other process is initially Byzantine. Also, a fault can arbitrarily change the $d$ and $f$ values of a Byzantine process. The fault transitions that affect a process, say $j$, of $\BA$ are as follows: (We include similar actions for $k$, $l$, and $g$)

\begin{tabbing}
\hspace*{1cm} \= $F1 \; :: \; \neg b.g \ad \neg b.j \ad \neg b.k \ad \neg b.l$ \hspace*{5mm} \= $\longrightarrow$ \hspace*{5mm} \= $b.j := \tru$\\
\> $F2 \; :: \; b.j$ \> $\longrightarrow$ \> $d.j, f.j := 0|1, \fals|\tru$
\end{tabbing}where $d.j := 0|1$ means that $d.j$ could be assigned either 0 or 1. In case of the general process, the second action does not change the value of any $f$-variable.

\noindent {\bf Goal of automated Addition of fault-tolerance. } \ 
Given the set of faults ($ F1 \& F2$), the goal of a synthesis algorithm is to start from the intolerant program ($BA_{\mathit{intol}_{j}}$) and generate the fault-tolerant program ($BA_{\mathit{tolerant}_{j}}$):
{
\small
\begin{tabbing}
\hspace*{.5cm}$BA_{tolerant_{j}}::$
\hspace*{.5cm} \= $\BA1_j \; :: \; (d.j = \bot) \; \ad \; (f.j = \fals) \; \ad \; (b.j = \fals)$ \hspace*{16mm} \= $\longrightarrow$ \hspace*{5mm} \= $d.j := d.g$\\
\> $\BA2_j \; :: \; (d.j \neq \bot) \; \ad \; (f.j = \fals)  \; \ad \; (d.j = d.l \; \vee \; d.j = d.k)$ \> $\longrightarrow$ \> $f.j := \tru$\\
\> $\BA3_j \; :: \; (d.l = 0 ) \; \ad \; (d.k = 0)  \; \ad \; (d.j = 1) \; \ad \; (f.j = 0)$ \> $\longrightarrow$ \> $d.j,f.j :=0, 0 |1$\\
\> $\BA4_j \; :: \; (d.l = 1 ) \; \ad \; (d.k = 1)  \; \ad \; (d.j = 0) \; \ad \; (f.j = 0)$ \> $\longrightarrow$ \> $d.j,f.j :=1, 0 |1$\\
\end{tabbing}
}

In the above program, the first action is identical to that of the intolerant program. The second action  is restricted to execute only in the states where another process has the same $d$ value. Actions ($ 3 \& 4 $) are for fixing the process decision through appropriate recovery.

%% file: readwrite.tex
\subsection{Group Computation: The Need for Modeling Read/Write Restrictions}
\label{sec:rw}

A process in a distributed program has a partial view of the program variables. For example, in the context of the Byzantine agreement example from Section \ref{sec:agreement}, process $j$ is allowed to read \linebreak $R_j = \{b.j, d.j, f.j, d.k, d.l, d.g\}$ and it is allowed to write $W_j = \{d.j, f.j\}$. Observe that this modeling prevents $j$ from knowing whether other processes are Byzantine. 

With such read/write restriction, if process $j$ were to include an action of the form `if $b.k$ is true then change $d.j$ to $0$' then it must also include a transition of the form `if $b.k$ is false then change $d.j$ to $0$'. In general, if transition $(s_0, s_1)$ is to be included as a transition of process $j$ then we must also include a corresponding equivalence class of transitions (called \textit{group} of transitions) that differ only in terms of variables that $j$ cannot read. The same mechanism has to be applied for removing transitions as well.

More generally, let $j$ be a process, let $R_j$ (respectively, $W_j$) be the set of variables that $j$ can read (respectively write), where $W_j \subseteq R_j$, and let $v_a(s_0)$ denote the value of variable $v_a$ in the state $s_0$. Then if $(s_0,s_1)$ is a transition that is included as a transition of $j$ then we must also include the corresponding equivalence class of transitions of the form ($s_2,s_3$) where $s_0$ and $s_2$ (respectively $s_1$ and $s_3$) are indistinguishable for $j$, i.e., they differ only in terms of the variables that $j$ cannot read. This equivalence class of transitions for $(s_0, s_1)$ is given by the following formula:

\begin{tabbing}
\noindent \hspace*{0mm} $\mathit{group}_j((s_0,s_1)) = \bigvee_{(s_2, s_3)}$\\
\hspace{3cm}$(\; $\= $\bigwedge_{v \not \in R_j} (v(s_0) = v(s_1) \;$ \= $\ad$ \; \=$v(s_2) = v(s_3)) \;\ad$\\
\noindent \> $\bigwedge_{v \in R_j} (v(s_0) = v(s_2)$ \> $\ad$ \> $v(s_1) = v(s_3)) \;)$.
\end{tabbing}

%% file: DL.tex
\subsection{Need for Deadlock Resolution}
\label{sec:dl}

During synthesis, we analyze the effect of faults on the given fault-intolerant program and identify a fault-tolerant program that meets the constraints of Problem Statement \probstmt. 
This involves addition of new transitions as well as removal of existing transitions. 
In this section, we utilize the Byzantine agreement problem to illustrate how deadlocks states get created during the execution of the synthesis algorithm and identify two general approaches for resolving them (be them sequential or parallel). 

\begin{itemize}
\item {\bf Deadlock scenario 1 and use of recovery actions. } \ 
One legitimate state, say $s$, for the Byzantine agreement program is a state where all processes are non-Byzantine, $d.g$ is $0$ and the decision of all non-generals is $\bot$. In this state, the general has chosen the value $0$ and no non-general has received any value. From this state, the general can become Byzantine and change its value from $0$ to $1$ arbitrarily. Hence, a non-general can receive either $0$ or $1$ from the general. Clearly, starting from $s$, in the presence of faults ($F1$ \& $F2$), the program ($BA_\mathit{intol}$) can reach a state, say $s_1$, where $d.g = d.j = d.k = 0, b.g = true, d.l = 1, f.l = 0$. From such a state, transitions of the fault-intolerant program violate agreement, if they allow $j$ (or $k$) and $l$ to finalize their decision. If we remove these safety violating transitions then there are no other transitions from state $s_1$. In other words, during synthesis, we encounter that state $s_1$ is a deadlock state. One can resolve this deadlock state by simply adding a \emph{recovery} transition that changes $d.l$ to $0$. 

\item {\bf Deadlock scenario 2 and need for elimination. } \
Again, consider the execution of the program ($BA_\mathit{intol}$) in the presence of faults ($F1$ \& $F2$) starting from state $s$ in the previous scenario. From $s$, the program can also reach a state, say $s_2$, where $d.g = d.j = d.k = 0, b.g = true, d.l = 1, f.l = 1$; state $s_2$ differs from $s_1$ in the previous scenario in terms of the value of $f.l$. Unlike $s_1$ in the previous scenario, since $l$ has finalized its decision, we cannot resolve $s_2$ by adding safe recovery. Since safe recovery from $s_2$ cannot be added, the only choice for designing a fault-tolerant program is to ensure that state $s_2$ is never reached in the fault-tolerant program by removing transitions that reach $s_2$ using backward reachability analysis. However, removal of such transitions can potentially create more deadlock states that have to be eliminated.

\end{itemize}

To maximize the success of synthesis algorithm, our approach to handle deadlock states is as follows: Whenever possible, we add recovery transition(s) from the deadlock states to a legitimate state. However, if no recovery transition(s) can be added from a deadlock state, we try to eliminate it by preventing the program from reaching the state. In this paper, we utilize parallelism to expedite these two aspects of deadlock resolution: adding recovery and eliminating deadlock states.

   \ifthink
\else
   \fi



%% file: FG.tex
\section{Approach 1: Parallelizing Group Computation}
\label{sec:FG}

In this section, we present our approach for parallelizing group computation to expedite synthesis of fault-tolerant programs. First, in Section \ref{sec:choices}, we identify different design choices in devising our parallel algorithm. Then, in Section \ref{sec:description}, we describe our approach for parallelizing the group computation. In Section \ref{sec:results}, we provide experimental results. Finally, in Section \ref{subsec:Time}, we analyze the experimental results to evaluate the effectiveness of parallelization for group computation.

\subsection{Design Choices}
\label{sec:choices}

The structure of the group computation permits an efficient way to parallelize it. In particular, whenever some recovery transitions are added for dealing with a deadlock state or some states are removed for ensuring that a deadlock state is not reached, we can utilize multiple threads in a master-slave fashion to expedite the group computation.The context of our approach targets multi-processor/core shared memory infrastructure. Although we did not specifically analyze the influence of local memory sharing on the performance, we expect our solution to give similar results when it uses multi-core or multi-processor architecture. During the analysis for utilizing multiple cores effectively, we make the following observations/design choices.

\begin{itemize}

\item {\bf Multiple BDD managers versus reentrant BDD package. } \
We chose to utilize different instances of BDD packages for each thread. Thus, at the time of group computation, each thread obtains a copy of the BDD corresponding to the recovery transitions being added. In part, this is motivated by the fact that existing parallel implementations have shown limited speedup (cf. Section \ref{sec:related}). Also, we argue that the increased space complexity of this approach is acceptable in the context of synthesis, since the time complexity of the synthesis algorithm is high (as opposed to model checking) and we often run out of time before we run out of space. 

\item {\bf Synchronization overhead. } \
The group computation is rather fine-grained, i.e., the time to compute a group of recovery transitions that are to be added to an input program is small (100-500ms on a normal machine). Hence, the overhead of creating multiple threads needs to be small. With this motivation, our algorithm creates the required set of threads up front and utilizes mutexes to synchronize them. This synchrnozation provides a significant benefit over creating and destroying threads for each group operation.

\item {\bf Load balancing. } \
Load balancing among several threads is desirable so that all threads take approximately the same amount of time in performing their task. To perform a group computation for recovery transitions being added, we need to evaluate the effect of read/write restrictions imposed by each process. A static way to parallelize this is to let each thread compute the set of transitions caused by read/write restrictions of a (given) subset of processes. A dynamic way is to consider the set of processes for which a group computation is to be performed as a {\em shared pool of tasks} and allow each thread to pick one task after it finishes the previous one. We find that given the small duration of each group computation, static partitioning of the group computation works better than dynamic partitioning since the overhead of dynamic partitioning is high. 
\end{itemize}

\subsection{Algorithm Description}
\label{sec:description}

Based on these design choices, the algorithm consists of three parts: initialization, assignment of tasks to worker threads and computation of group with worker threads. 

\noindent {\bf Initialization. } \
In the initialization phase, the master thread creates all required worker threads by calling the algorithm {\sf InitiateThreads} (cf. Algorithm \ref{alg:gth12}). These threads stay idle until a group computation is required and terminate when the synthesis algorithm ends. Due to the design choice for load balancing, the algorithm distributes the work load among the available threads statically (Lines 4-8). Then, it creates all the required worker threads (Line 10).

\floatname{algorithm}{Algorithm}
\begin{algorithm}
\caption{InitiateThreads}
\label{alg:gth12}
\algsetup{linenosize = \scriptsize, indent = 1cm}
\scriptsize{
\begin{algorithmic}[1]
\INPUT{  $\mathit{noOfProcesses}$,  $\mathit{noOfThreads}$.}
\ \\ 
\ \\
\IF {$\mathit{noOfProcesses}   \; < \; \mathit{noOfThreads}$} 
\RETURN ERROR;
\ENDIF
\FOR {$i := 0$ \textbf{to} $\mathit{noOfThreads}-1 $}
\STATE	 $\mathit{BDDMgr}[i]$  = $\mathit{Clone(masterBDDManager)}$ ;

\STATE $\mathit{startP}[i]$ := $\lfloor \frac{i \;\; \times \;\; \mathit{noOfProcesses}}  { \mathit{noOfThreads} } \rfloor $;
\STATE $\mathit{endP}[i]$ := $\lfloor \frac{(i+1) \;\; \times \;\; \mathit{noOfProcesses}}  { \mathit{noOfThreads} } \rfloor -1$;
\ENDFOR
\FOR {$\mathit{thID}$ := $0$ \textbf{to} $\mathit{noOfThreads}-1 $}
\STATE $\mathit{SpawnThread}$  $\rightsquigarrow$  $\mathit{WorkerThread}(\mathit{thID})$;
\ENDFOR
\end{algorithmic}
}
\end{algorithm}

\mbox{}

\noindent {\bf Tasks for worker thread. } \
Initially, the algorithm {\sf WorkerThread} (cf. Algorithm \ref{alg:gth13}) locks the mutexes $\mathit{mutexStart}$ and $\mathit{mutexStop}$ (Lines 1-2). Then, it waits until the master thread unlocks the $\mathit{mutexStart}$ mutex (Line 5). At this point, the worker thread starts computing the part of the group associated with this thread. This section of {\sf WorkerThread} (Lines 7-15) is similar to the computing groups in the sequential setting except rather than finding the group for all the processes, the {\sf WorkerThread} algorithm finds the group for a subset of processes (Line 8). The function $\mathit{AllowWrite}$ relaxes a predicate with respect to the variables that the corresponding process is allowed to modify. The function $\mathit{Transfer}$ transfers a BDD from one manager to another manager. And, the function $\mathit{FindGroup}$ adds read restrictions to a group predicate. When the computation is completed, the worker thread notifies the master thread by unlocking the mutex $\mathit{mutexStop}$ (Line 17).

\floatname{algorithm}{Algorithm}
\begin{algorithm}
\caption{WorkerThread}
\label{alg:gth13}
\algsetup{linenosize = \scriptsize, indent = 1cm}
\scriptsize{
\begin{algorithmic}[1]
\INPUT{  $\mathit{thID} $.}
 \ \\
 \ \\
 \emph{// Initial locking of the mutexes} \\

\STATE $\mathit{mutex\_lock(thData}[\mathit{thID}].\mathit{mutexStart})$;
\STATE $\mathit{mutex\_lock(thData[thID].mutexStop})$;

\WHILE{$\tru$}
\STATE  \emph{// Waiting for signal from the master thread}\\
\STATE $\mathit{mutex\_lock}(\mathit{thData}[\mathit{thID}].\mathit{mutexStart})$;
\STATE $\gtr[\id]$ :=  $\fals$;
\STATE $\mathit{tPred}$ := $\mathit{endP}[\mathit{thID}] - \mathit{startP}[\mathit{thID}]+1 $  ;
\FOR {$i := 0$ \textbf{to}  $(\mathit{endP}[\mathit{thID}] -\mathit{startP}[\mathit{thID}])+ 1$}

\STATE	 $\mathit{tPred}[i]$ := $\mathit{thData}[\mathit{thID}].\mathit{trans} \;\;\ad \;\; \mathit{allowWrite}[i+ \mathit{startP}[\mathit{thID}]].\mathit{Transfer}(\mathit{BDDMgr}[\mathit{thID}])$; \\
\STATE      $\mathit{tPred}[i]$ := $\mathit{FindGroup}(\mathit{tPred}[i],  i, \mathit{thID})$;
\ENDFOR
\STATE  $\mathit{thData}[\mathit{thID}].\mathit{result}$ := $\fals$;

\FOR {$i := 0$ \textbf{to} $ (\mathit{endP}[\mathit{thID}] -\mathit{startP}[\mathit{thID}])+ 1$}

\STATE  $\mathit{thData}[\mathit{thID}].\mathit{result}$ :=  $\mathit{thData}[\mathit{thID}].\mathit{result}$ $\vee$ $\mathit{tPred}[i]$;

\ENDFOR

\STATE  \emph{// Triggering the master thread that this thread is done}\\
\STATE $\mathit{mutex\_unlock}(\mathit{thData}[\mathit{thID}].\mathit{mutexStop})$;
\ENDWHILE

\end{algorithmic}
}
\end{algorithm}

\noindent {\bf Tasks for master thread. } \
Given transition set $\mathit{tr}$, the master thread copies $\mathit{tr}$ to each instance of the BDD package used by the worker threads (cf. Algorithm \ref{alg:gth}, Lines 3-5). Then it assigns a subset of group computation to the worker threads (Lines 6-8) and unlocks them. After the worker threads complete, the master thread collects the results and returns the group BDD associated with the input $\mathit{tr}$.

\floatname{algorithm}{Algorithm}
\begin{algorithm}
\caption{MasterThread}
\label{alg:gth}
\algsetup{linenosize = \scriptsize, indent = 1cm}
\scriptsize{
\begin{algorithmic}[1]
\INPUT{transitions set  $\mathit{thisTr}$.}
\ \\
\OUTPUT{transition group $\gAll$.}
\ \\
\ \\

\STATE $\tr$ :=  $\mathit{thisTr}$;
\STATE $\gAll$ :=  $\fals$;
\FOR {$i := 0$ \textbf{to} $\mathit{NoOfThreads} - 1$}
\STATE	 $\mathit{threadData}[i].\mathit{trans}$ := $\mathit{trans.Transfer}(\mathit{BDDMgr}[\mathit{thID}])$;
\ENDFOR

\emph{//  all idle threads to start computing the group}\\
\FOR {$i := 0$ \textbf{to} $\mathit{NoOfThreads} -1 $}
\STATE	 $\mathit{mutex\_unlock}(\mathit{thData}[i].\mathit{mutexStart})$;
\ENDFOR
\ \\
\emph{// Waiting for all threads to finish computing the group}\\
\FOR {$i := 0$ \textbf{to} $\mathit{NoOfThreads} - 1$}
\STATE	 $\mathit{mutex\_lock}(\mathit{thData}[i].\mathit{mutexStop})$;
\ENDFOR
\ \\
\emph{// Merging the results from all threads}\\
\FOR {$i := 0$ \textbf{to} $\mathit{NoOfThreads}-1 $}
\STATE	 $\gAll$ := $\gAll$ + $\mathit{thData}[i].\mathit{results}$;
\ENDFOR
\ \\
\RETURN  $\gAll$;

\end{algorithmic}
}
\end{algorithm}

%% file: results.tex
\subsection{Experimental Results}
\label{sec:results}


In this section, we describe the respective experimental results in the context of the Byzantine agreement (described in Section \ref{sec:agreement}). Throughout this section, all experiments are run on a Sun Fire V40z
with  4 dual-core Opteron processors and 16 GB RAM. The BDD representation of the Boolean formulae has been done using the C++ interface to the CUDD package developed at University of Colorado \cite{cudd}. Throughout this section, we refer to the original implementation of the synthesis algorithm (without parallelism) as \emph{sequential} implementation. We use \emph{X threads} to refer to the parallel algorithm that utilizes $X$ threads.

We would like to note that the synthesis time duration differs between the sequential implementation in this paper and the one in \cite{bk07} due to other unrelated improvements on the sequential implementation itself. However, the sequential, and the parallel implementations differ only in terms of the modification described in Section \ref{sec:description}.

We note that our algorithm is deterministic and the testbed is dedicated. Hence, the only non-deterministic factor in time for synthesis is synchronization among threads. Based on our observations and experience, this factor has a negligible impact and, hence, multiple runs on the same data essentially reproduce the same results.

\begin{figure}[htp]
     \centering
     \subfigure[Deadlock Resolution Time]{
          \label{fig:drtsaBA}
          \includegraphics[width=.45\textwidth]{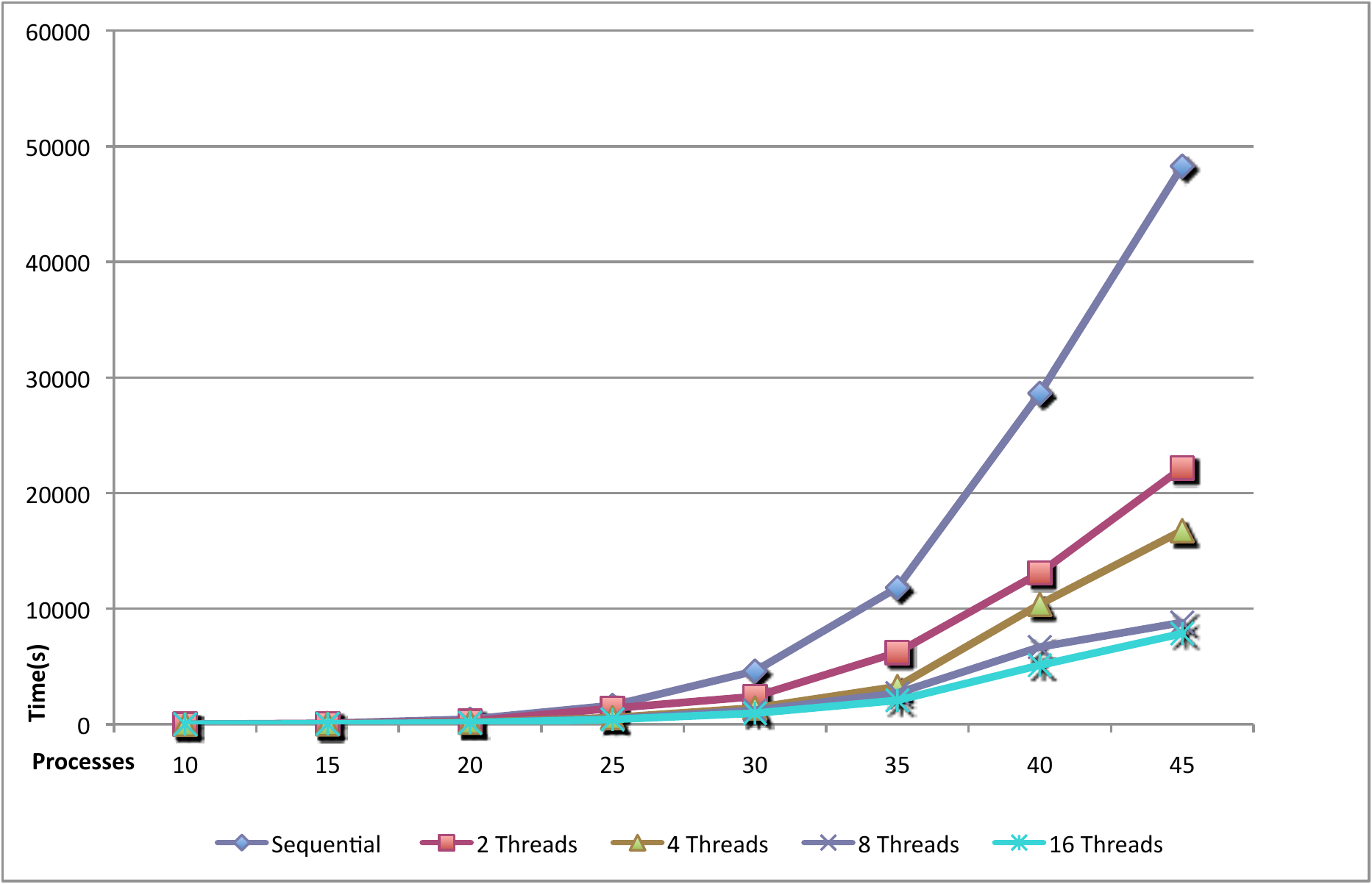}}
     \hspace{.1in}
     \subfigure[Total Synthesis Time ]{
          \label{fig:sttsbBA}
          \includegraphics[width=.45\textwidth]{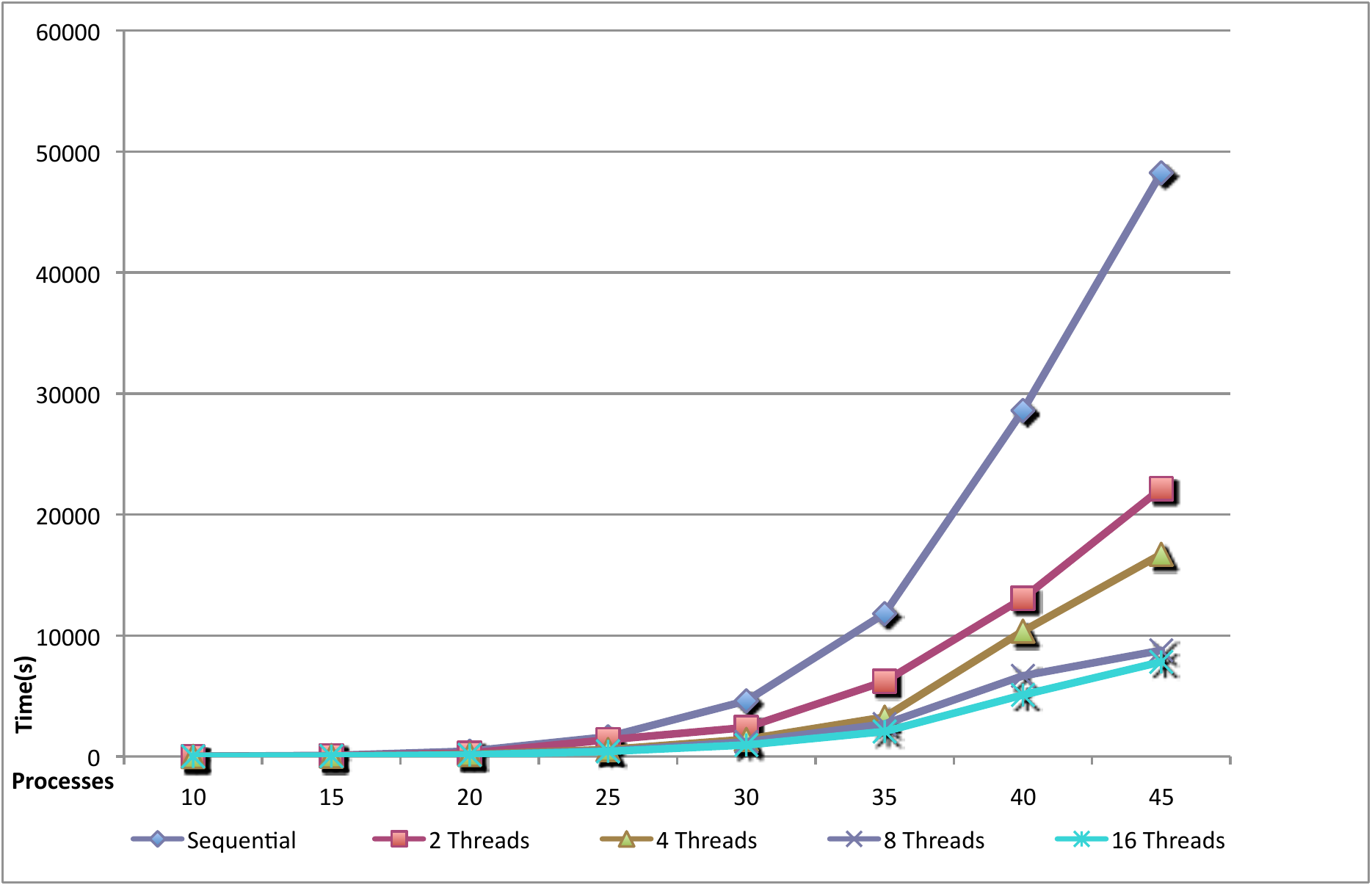}}\\
     \caption{The time required to (a) resolve deadlock states and (b) to synthesize a fault-tolerant program for several numbers of non-general processes of $\BA$ using sequential and parallel algorithms. The BA has a state space  $\approx  4*10^{1.08 x}$  and reachable state space  $\geq 2 * 10 ^ {0.78 x}$ where $x$ is the number of process. }    
 \label{fig:tsBA}
\end{figure}

In Figure \ref{fig:tsBA}, we show the results of using the sequential approach versus the parallel approach (with multiple threads) to perform the synthesis.  All the tests have shown that we gain a significant speedup. For example, in the case  of $45$ non-general processes and $8$ threads we gain a speedup of $6.1$~. We can clearly see that the parallel $16$-thread version is faster than the corresponding 8-threads version. This was surprising given that there are only 8 cores available. However, upon closer observation, we find that the group computation that is parallelized using threads is fine-grained. Hence, when the master thread uses multiple slave threads for performing the group computation, the slave threads complete quickly and therefore cannot utilize the available resources to the full extent. Hence, creating more threads (than available processors) can improve the performance further.

%% file: analyses.tex
\subsection{Group Time Analysis }
\label{subsec:Time}

In this section, we focus on the effectiveness of the parallelization of group computation by considering the time taken for it in sequential and parallel implementation. Towards this end, we analyze the group computation time for sequential and parallel implementations in the context of three examples: Byzantine agreement, agreement in the presence of failstop and Byzantine faults, and token ring \cite{ak981}. The results for these examples are included in Tables \ref{tab:groupdataByz}-\ref{tab:groupdataToken}. The number of cores used is equal to the number of threads.

To understand the speedup gain provided by our algorithm in Section \ref{sec:results}, we evaluated the experimental results closely. As an example, consider the case of $32$ $BA$ processes. For sequential implementation, the total synthesis time is $59.7$ minutes of which $55$ are used for group computation. Hence, the ideal completion time with 4 cores is 18.45 minutes ($55/4 + 4.7$). By comparison, the actual time taken in our experiment was $19.1$ minutes. Thus, the speedup gained using this approach is close to the ideal speedup. 

In some cases, the speedup ratio is less than the number of threads. This is caused by the fact that each group computation takes a very small time and incurs an overhead for thread synchronization. Moreover, as mentioned in Section \ref{sec:dl}, due to the overhead of load balancing, we allocate tasks of each thread statically. Thus, the load of different threads can be slightly uneven.
We also observe that the speedup ratio increases with the number of processes in the program being synthesized. This implies that the parallel algorithm will scale to larger problem instances. 

\begin{table}[h]
\begin{center}
\tiny
\begin{tabular}{|c|| c|| c|| c|c|| c|c|| c|c|  }
\hline
&  & {\emph{Sequential}}& \multicolumn{2}{c||}{\emph{2-threads}} & \multicolumn{2}{c||}{\emph{4-threads}} & \multicolumn{2}{c|}{\emph{8-threads}} \\
\cline{3-9}

&  &  &   & &   & &  &   \\[1pt]
\textbf{No. of}& \textbf{Reachable}  & \textbf{Group}& \textbf{Group} & \textbf{Speedup} & \textbf{Group} & \textbf{Speedup}& \textbf{Group} & \textbf{Speedup}\\[1pt]
\textbf{Processes}& \textbf{States}& \textbf{Time}& \textbf{Time} & \textbf{Ratio} & \textbf{Time} & \textbf{Ratio}& \textbf{Time} & \textbf{Ratio}\\[1pt]
\hline\hline
&  &  &   & &   & &  &   \\[1pt]

$15$	&$10^{11}$   &50			&29 &1.72	&17 &2.94	&11& 4.55	  \\[2pt]

$24$	&$10^{17}$   &652		&346 &1.88	&185 &3.52	&122& 5.34	 \\[2pt]

$32$	&$10^{22}$   &3347		&1532 &2.18	&848 &3.95	&490& 6.83	  \\[2pt]

$48$	&$10^{33}$   &33454		&14421 &2.32	&7271 &4.60	&3837& 8.72	\\[2pt]


\hline
\end{tabular}
\end{center}
\caption{\small Group computation time for Byzantine Agreement.}
\label{tab:groupdataByz}
\end{table}

\begin{table}[h]
\begin{center}
\tiny
\begin{tabular}{|c|| c|| c|| c|c|| c|c|| c|c|  }
\hline
&  & {\emph{Sequential}}& \multicolumn{2}{c||}{\emph{2-threads}} & \multicolumn{2}{c||}{\emph{4-threads}} & \multicolumn{2}{c|}{\emph{8-threads}} \\
\cline{3-9}

&  &  &   & &   & &  &   \\[1pt]
\textbf{No. of}& \textbf{Reachable}  & \textbf{Group}& \textbf{Group} & \textbf{Speedup} & \textbf{Group} & \textbf{Speedup}& \textbf{Group} & \textbf{Speedup}\\[1pt]
\textbf{Processes}& \textbf{States}& \textbf{Time}& \textbf{Time} & \textbf{Ratio} & \textbf{Time} & \textbf{Ratio}& \textbf{Time} & \textbf{Ratio}\\[1pt]
\hline\hline
&  &  &   & &   & &  &   \\[1pt]

$10$	&$10^{10}$   &53			&24& 2.21		&23 &2.30 	&30&1.77 	  \\[2pt]
$15$	&$10^{15}$   &624			&319 & 1.96		&175 &3.57 	&174& 3.59 	  \\[2pt]
$20$	&$10^{20}$   &4473			&2644 &1.69 		&1275 & 3.51	&1128&3.97  	 \\[2pt]
$25$	&$10^{25}$   &26154 		&11739&2.23		&6527  &4.01 		&5692	& 4.59		  \\[2pt]


\hline
\end{tabular}
\end{center}
\caption{\small Group computation time for the Agreement problem in the presence of failstop and Byzantine faults.}
\label{tab:groupdataByzFailstop}
\end{table}

\begin{table}[h]
\begin{center}
\tiny
\begin{tabular}{|c|| c|| c|| c|c|| c|c|| c|c|  }
\hline
&  & {\emph{Sequential}}& \multicolumn{2}{c||}{\emph{2-threads}} & \multicolumn{2}{c||}{\emph{4-threads}} & \multicolumn{2}{c|}{\emph{8-threads}} \\
\cline{3-9}

&  &  &   & &   & &  &   \\[1pt]
\textbf{No. of}& \textbf{Reachable} & \textbf{Group}& \textbf{Group} & \textbf{Speedup} & \textbf{Group} & \textbf{Speedup}& \textbf{Group} & \textbf{Speedup}\\[1pt]
\textbf{Processes}& \textbf{States} & \textbf{Time}& \textbf{Time} & \textbf{Ratio} & \textbf{Time} & \textbf{Ratio}& \textbf{Time} & \textbf{Ratio}\\[1pt]
\hline\hline
&  &  &   & &   & &  &   \\[1pt]

$30$	&$10^{14}$   &0.32		&0.15 &2.12	&0.10 &3.34	&0.12& 2.75	  \\[2pt]

$40$	&$10^{19}$   &0.84		&0.36 &2.34	&0.22 &3.84	&0.23& 3.59	 \\[2pt]

$50$	&$10^{23}$   &1.82		&0.68 &2.68	&0.39 &4.66	&0.42& 4.37	  \\[2pt]

$60$	&$10^{28}$   &3.22		&1.22 &2.63	&0.67 &4.80	&0.64& 5.01	\\[2pt]

$70$	&$10^{33}$   &5.36		&1.91 &2.80	&1.06 &5.05	&0.86& 6.23	\\[2pt]

$80$	&$10^{38}$   &7.77		&2.94 &2.64	&1.53 &5.09	&1.23& 6.30	\\[2pt]


\hline
\end{tabular}
\end{center}
\caption{\small Group computation time for  token ring.}
\label{tab:groupdataToken}
\end{table}


An interesting as well as surprising observation is that when the state space is large enough then the speedup ratio is more than the number of threads. This behavior is caused by the fact that with parallelization, each thread is working on smaller BDDs during the group computation. To understand this behavior, we conducted experiments where we created the threads to perform the group computation and forced them to execute sequentially by adding extra synchronization. We found that such pseudo-sequential run took less time than that used by a purely sequential run. 

%% file: CGnew.tex
\section{Approach 2: Alternative (Conventional) Approach} 
\label{sec:cgp}

A traditional approach for parallelization in the context of resolving deadlock states, say $ds$, would be to partition the deadlock states into multiple threads and allow each thread to handle the partition assigned to it. For example, we can partition $\ds$ using the partition predicates, $\pp_i, 1 \leq i \leq n$, such that $\bigvee_{i=1}^n (\pp_i \ad \ds) = \ds$. Thus, if two threads are available during synthesis of the Byzantine agreement program then we can let $\pp_1 = (d.j = 0)$ and $\pp_2 = (d.j \neq 0)$.

Next, in Section \ref{sec:CGdesign}, we discuss some of the design choices we considered for this approach. Subsequently, we describe experimental results in Section \ref{sec:CGexperimental}. We argue that for such an approach to work in synthesizing distributed programs, group computation must itself be parallelized. 

\subsection{Design Choices}
\label{sec:CGdesign}

To efficiently partition deadlock states among threads, one needs to design a method such that (1) deadlock states are evenly distributed among worker threads, and (2) states considered by different threads for elimination have a small overlap during backtracking. Regarding the first constraint, we can partition deadlock states based on values of some variable and evaluate the size of corresponding BDDs by the number of minterms that satisfy the corresponding formula. Regarding the second constraint, we expect that the overhead for such a split is as high as it requires detailed analysis of program transitions. Hence, instead of satisfying this constraint, we choose to add limited synchronization among threads so that the overlap in the explored states by different threads is small.

After partitioning, one thread would work independently as long as it does not affect states visited by other threads. As discussed in Section \ref{sec:dl}, to resolve a deadlock state, each thread explores a part of the state space using backward reachability. Clearly, when states visited by two threads overlap, we have two options: (1) perform synchronization so that only one thread explores any state or (2) allow two threads to explore the states concurrently and resolve any inconsistencies that may be created. 

We find that the first option by itself is very expensive/impossible due to the fact that with the use of BDDs, each thread explores a set of states specified by the BDD. And, since each thread begins with a set of deadlock states and performs backward reachability, there is a significant overlap among states explored by different threads. Hence, the first option is likely to essentially reduce the parallel run to a sequential run. For this reason, we focus on the second approach where each thread explored the states concurrently. (We also use some heuristic based synchronization where we maintained a set of $\mathit{visited}$ states that each thread checked before performing backward state exploration. This technique provided only a small performance benefit.)

\begin{figure}[h]
   \centering
   \includegraphics[scale= 0.7 ]{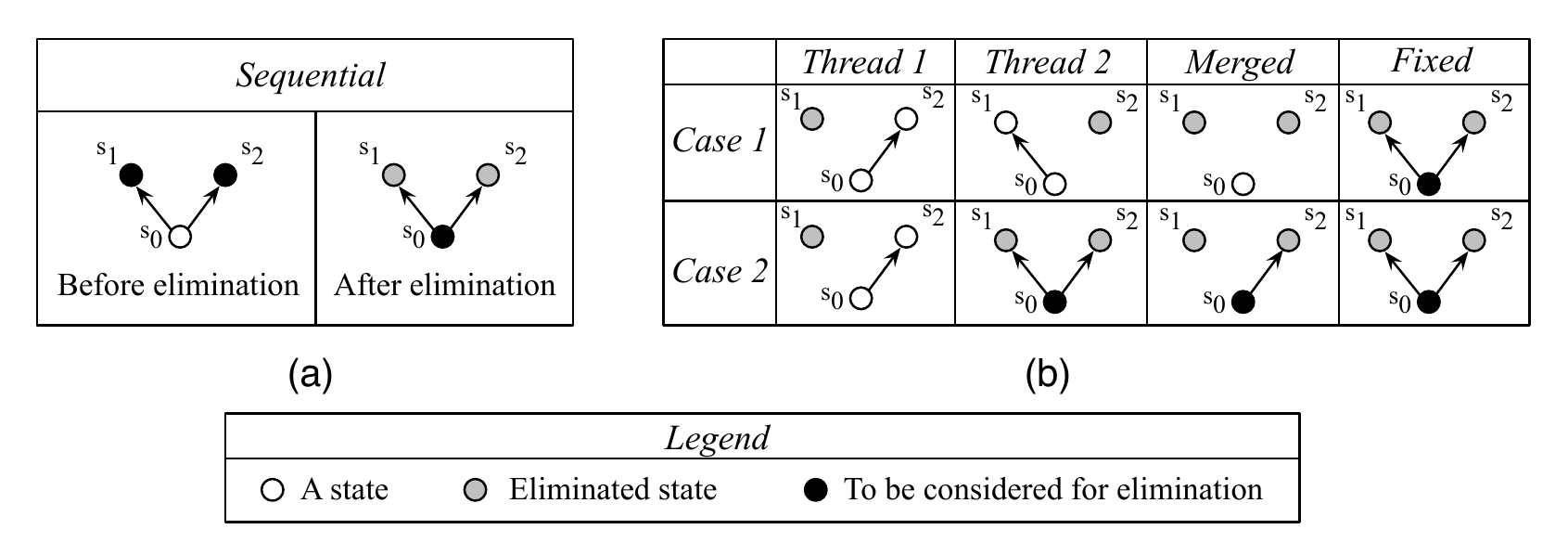}
   \caption{Inconsistencies raised by concurrency.}
   \label{fig:cases}
\end{figure}

\noindent {\bf Inconsistency Resolution. } \
When threads explore states concurrently, some inconsistencies may be created. Next, we give a brief overview of the inconsistencies that may occur due to concurrent state exploration and manipulation by different threads and identify how we can resolve them. Towards this end, let $s_1$ and $s_2$ be two states that are considered for deadlock elimination and $(s_0, s_1)$ and $(s_0, s_2)$ be two program transitions for some $s_0$. A sequential elimination algorithm, removes transitions $(s_0, s_1)$ and $(s_0, s_2)$ which causes $s_0$ to be a new deadlock state (cf. Figure \ref{fig:cases}.a). This in turn requires that state $s_0$ itself must be made unreachable. If $s_0$ is unreachable then including the transition $(s_0, s_1)$ in the synthesized program is harmless. In fact, it is desirable since including this transition also causes other transitions in the corresponding group to be included as well. And, these grouped transitions might be useful in providing recovery from other states. 
Hence, it puts $(s_0, s_1)$ and $(s_0, s_2)$ (and corresponding group) back into the program being synthesized and it continues to eliminate the state $s_0$. However, when multiple worker threads, say $\thr_1$ and $\thr_2$, run concurrently, some inconsistencies may be created. We describe some of these inconsistencies and our approach to resolve them next.

\noindent \textbf{Case 1. } \ States $s_1$ and $s_2$ are in different partitions. Hence, $\thr_1$ eliminates $s_1$ which in turn removes the transition $(s_0, s_1)$, and, $\thr_2$ eliminates  $s_2$ which removes the transition $(s_0, s_2)$ (cf. Figure \ref{fig:cases}.b). Since each thread works on its own copy, neither thread tries to eliminate $s_0$, as they do not identify $s_0$ as a deadlock state. Subsequently, when the master thread merges the results returned by $\thr_1$ and $\thr_2$, $s_0$ becomes a new deadlock state which has to be eliminated while the group predicates of transitions $(s_0, s_1)$ and $(s_0, s_2)$ have been removed unnecessarily. In order to resolve this case, we re-introduce all outgoing transitions that start from $s_0$ and mark $s_0$ as a state that has to be eliminated in subsequent iterations.

\noindent \textbf{Case 2. } \ Due to backtracking behavior of the elimination algorithm, it is possible that $\thr_1$ and $\thr_2$ consider common states for elimination. In particular, if $\thr_1$ considers $s_1$ and $\thr_2$ considers both $s_1$ and $s_2$ for elimination (cf. Figure \ref{fig:cases}.b), after merging the results, no new deadlock states are introduced. However, $(s_0, s_1)$ would be removed unnecessarily. In order to resolve this case, we collect all the states that worker threads failed to eliminate and replace all incoming transitions into those states.

\subsection{Experimental Results }
\label{sec:CGexperimental}
 
We also implemented this approach for parallelization. The results for the problem of Byzantine agreement are as shown in Table \ref{tbl:Elim}.
From these results, we notice that the improvement in the performance was small. 


\begin{table}[h]
\begin{center}
\tiny
\begin{tabular}{|c|| c|| c| c|| c| c|  }
\hline
&  & \multicolumn{2}{c||}  {} &  \multicolumn{2}{c|}    {}    \\[1pt]

&  &\multicolumn{2}{c||}  {\emph{Sequential}} &\multicolumn{2}{c|}{\emph{Parallel Elimination with 2-threads}} \\

\cline{3-6}
&  &  & & &       \\[1pt]

\textbf{No. of}& \textbf{Reachable} & \textbf{Deadlock}& \textbf{Total} & \textbf{Deadlock} & \textbf{Total}\\[1pt]
\textbf{Processes}& \textbf{States} & \textbf{Resolution Time}& \textbf{Synthesis Time} & \textbf{Resolution Time} & \textbf{Synthesis Time}\\[1pt]
\hline\hline
&  &  & & &       \\[1pt]

$10$	&$10^{7}$   &7		&9          &8		&9 	  	\\[2pt]
$15$	&$10^{12}$   &78		&85        &78		& 87	 	\\[2pt]
$20$	&$10^{14}$   &406		&442      &374		& 417	\\[2pt]
$25$	&$10^{18}$   &1,503	&1,632   &1,394	& 1,503	\\[2pt]
$30$	&$10^{21}$   &4,302	&4,606   &3,274	& 3,518	\\[2pt]
$35$	&$10^{25}$   &11,088	&11,821 &10,995	& 11,608	\\[2pt]
$40$	&$10^{28}$   &27,115	&28,628 &21,997	& 23,101	\\[2pt]
$45$	&$10^{32}$   &45,850	&48,283 &39,645	& 41,548	\\[2pt]


\hline
\end{tabular}
\end{center}
\caption{\small The time required to synthesis tolerant program for several numbers of non-general processes of $BA$ in sequential and by partitioning deadlock states using parallelism.}
     \label{tbl:Elim}
\end{table}

%% file: related.tex
\section{Related Work}
\label{sec:related}

Automated program synthesis and revision has been studied from various perspectives. Inspired by the seminal work by Emerson and Clarke \cite{ec82}, Arora, Attie, and Emerson \cite{aae98} propose an algorithm for synthesizing fault-tolerant programs from CTL specifications. Their method, however, does not address the issue of the addition of fault-tolerance to existing programs. Kulkarni and Arora \cite{ka00} introduce enumerative synthesis algorithms for automated addition of fault-tolerance to centralized and distributed programs. In particular, they show that the problem of adding fault-tolerance to distributed programs is NP-complete. In order to remedy the NP-hardness of the synthesis of fault-tolerant distributed programs and overcome the state explosion problem, we proposed a set of symbolic heuristics \cite{bk07}, which allowed us to synthesize programs with a state space size of $10^{30}$ and beyond.

Ebnenasir \cite{e07} presents a divide-and-conquer method for synthesizing \emph{failsafe} fault-tolerant distributed programs. A failsafe program is one that does not need to satisfy its liveness specification in the presence of faults. Thus, a respective synthesis algorithm does not need to resolve deadlock states outside the invariant predicate. Moreover, Ebnenasir's synthesis method resolves deadlock states inside the invariant predicate in a sequential manner.

We have also presented an approach \cite{fs094} for utilizing multi-core technology in the design of self-stabilizing programs, i.e., a program that ensures that starting from an arbitrary state, it recovers to a legitimate state. This work utilizes parallelization of group computation as well as another approach for expediting the design of stabilizing programs. However, due to the nature of the problem involved, parallelization of group computation is more effective in deadlock resolution than in design of stabilizing programs \cite{fs094}.

Parallelization of symbolic reachability analysis has been studied in the model checking community from different perspectives. In \cite{el07, elc07, els06}, the authors propose solutions and analyze different approaches to parallelization of the \emph{saturation}-based generation of state space in model checking. In particular, in \cite{elc07}, the authors show that in order to gain speedups in saturation-based parallel symbolic verification, one has to pay a penalty for memory usage of up to 10 times, that of the sequential algorithm. Other efforts range from simple approaches that essentially implement BDDs as two-tiered hash tables \cite{mh98, sb96}, to sophisticated approaches relying on \emph{slicing} BDDs \cite{ghs05} and techniques for \emph{workstealing} \cite{ghis05}. However, the resulting implementations show only limited speedups.

%% file: concl.tex
\section{Conclusion }
\label{sec:concl}
{\bf Summary. } \
In this paper, we focused on improving the synthesis of fault-tolerant programs from their fault-intolerant version. We focused on two approaches for expediting the performance of the synthesis algorithm by using multi-core computing. We showed that the approach of partitioning deadlock states provides a small improvement. And,  the approach based on parallelizing the group computation -- that is caused by distribution constraints of the program being synthesized-- provides a significant benefit that is close to the ideal, i.e., equal to the number of threads used. Moreover, the performance analysis shows that this approach is scalable in that if more cores were available, our approach can utilize them effectively. 

\noindent {\bf Lessons Learnt. } \
As shown in \cite{bk07}, there are two main bottlenecks in synthesizing fault-tolerant programs: {\em generation of fault-span} which is essentially a reachability problem that has been studied extensively in the context of model checking and {\em deadlock resolution} that corresponds to adding recovery paths from states reached in the presence of faults. The results in this paper show that a traditional approach (Section \ref{sec:cgp}) of partitioning deadlock states provides a small improvement. However, it helped identify an alternative approach for parallelization that is based on the distribution constraints imposed on the program being synthesized. 

The performance improvement with the use of the distribution constraints is significant. In fact, for most cases, the performance was close to the ideal speedup. What this suggests is that for the task of deadlock resolution, a simple approach based on parallelizing the group computation (as opposed to a reentrant BDD package that permits multiple concurrent threads or partition of deadlock states etc.) that is caused due to distribution constraints will provide the biggest benefit in performance. Moreover, the group computation itself occurs in every aspect of synthesis where new transitions have to be added for recovery or existing transitions have to be removed for preventing safety violation or breaking cycles that prevent recovery to the invariant. 
Hence, the approach of parallelizing the group computation will be effective in the synthesis of distributed programs. 

\noindent {\bf Impact. } \
Automated synthesis has been widely believed to be significantly more complex than automated verification. When we evaluate the complexity of automated synthesis of fault-tolerance, we find that it fundamentally include two parts: (1) analyzing the existing program and (2) transforming it to ensure that it meets the fault-tolerance properties. The first part closely resembles with program verification and techniques for efficient verification are directly applicable to it. What this paper shows is that the complexity of the second part can be significantly remedied by the use of parallelization in a {\em simple} and {\em scalable} fashion. Moreover, if we evaluate the typical inexpensive technology that is currently being used or is likely to be available in {\em near future}, it is expected to be 2-16 core computers. And, the first approach used in this paper is expected to be the most suitable one for utilizing these multicore computers to the fullest extent.  Also, since the group computation is caused by distribution constraints of the program being synthesized, as discussed in Section \ref{sec:cgp}, it is guaranteed to be required even with other techniques for expediting automated synthesis. For example, it can be used in conjunction with the approach in Section \ref{sec:cgp} as well as the approach that utilizes symmetry among processes being synthesized.